\documentclass[iop]{emulateapj}
\usepackage{bm}
\usepackage{multirow}
\usepackage{color}
\usepackage{float}
\usepackage{soul}
\usepackage{natbib}
\usepackage{amsmath}
\usepackage{amsfonts}
\usepackage{amssymb}
\usepackage{epsfig}
\usepackage{hyperref}
\usepackage{graphicx}
\usepackage{booktabs}

\newcommand{\beq}{\begin{equation}}
\newcommand{\eeq}{\end{equation}}
\newcommand{\beqa}{\begin{eqnarray}}
\newcommand{\eeqa}{\end{eqnarray}}

\newcommand{\jcap}{J. Cosmol. Astropart. Phys.}

\begin{document}

\title{Halo Pressure Profile through the Skew Cross-Power Spectrum of
Sunyaev-Zel'dovich Effect and CMB Lensing in {\it Planck}}
\author{Nicholas Timmons$^{1}$, Asantha Cooray$^{1}$, Chang Feng$^{1}$, Brian Keating$^{2}$}

\affiliation{$^{1}$Department of Physics  and  Astronomy, University  of California, Irvine, CA 92697}
\affiliation{$^{2}$Department of Physics, University of California, San Diego, La Jolla, CA 92093}

\begin{abstract}

We measure the Cosmic Microwave Background (CMB) skewness power spectrum in {\it Planck}, using  frequency maps of the HFI instrument and the Sunyaev-Zel'dovich (SZ) component map. The two-to-one skewness power spectrum measures the cross-correlation between CMB lensing and the thermal SZ effect. We also directly measure the same cross-correlation using {\it Planck} CMB lensing map and the SZ map and compare it to the cross-correlation derived from the skewness power spectrum. We model fit the SZ power spectrum and CMB lensing-SZ cross power spectrum via the skewness power spectrum to constrain the gas pressure profile of dark matter halos. The gas pressure profile is compared to existing measurements in the literature including a direct estimate based on the stacking of SZ clusters in {\it Planck}.

\end{abstract}

\maketitle

\section{Introduction}
The importance of the {\it Planck} mission \citep{Planck2011} to cosmology can not be overstated. The measurement of cosmic microwave background (CMB) anisotropies has allowed for increased accuracy in measurements of cosmological parameters. Going beyond primary anisotropies {\it Planck} frequency maps can be used to probe higher order correlations that arise from secondary effects like the Sunyaev-Zel'dovich (SZ) effect \citep{SZ1980}, integrated Sachs-Wolfe effect (ISW) \citep{SW1967} and gravitational lensing \citep{Smith2007} among others.

The SZ effect is the result of inverse Compton scattering of photons off free electrons. The SZ effect on the cosmic microwave background radiation (CMB) is the result of gas being heated from the pressure inside galaxy clusters. Measurement of the SZ effect is a key tracer of the hot electron gas in the intra-cluster medium (ICM). Due to its unique frequency spectrum relative to CMB black-body the thermal SZ effect can be separated in multi-frequency CMB maps \citep{Cooray2000}. Gravitational lensing of the CMB is caused by the intervening mass distribution and is a tracer of the line of sight projected dark matter potential. The integrated lensing map can be extracted from a quadratic \citep{Okamoto2003} and likelihood \citep{Hirata2008} estimators on CMB.  A non-zero correlation between the lensing effect and the SZ effect establishes the relationship between dark matter and hot intra-cluster gas \citep{Hill2014}. This is an excellent probe of the large scale distribution of gas. Several studies have attempted to look at this correlation (\citealt{Cooray2000,Hill2014}) including one using the WMAP data \citep{Calabrese2010}. Here we update the previous work by incorporating data from {\it Planck}.  

The cross-correlation between CMB lensing and thermal SZ results in a non-Gaussian signal at the three-point level of the correlation function (\citealt{Spergel1999,Cooray2000}). While challenging to measure directly the bispectrum can be collapsed into a sum of two-point functions in what is known as the skewness power spectrum involving a squared temperature-temperature correlation. As has been shown in previous work (\citealt{Cooray2001,Munshi2011}) the skewness spectrum, related to the CMB-CMB lensing-SZ bispectrum, can be probed through the cross-correlation of a temperature squared map and a map of the SZ effect. This skewness spectrum contains all the information from the bispectrum once the estimator is appropriately weighted. The three point correlation function using only the CMB is an independent look at the lensing-SZ cross-correlation.

The amplitude of the non-Gaussian signals arising from the SZ effect can help constrain physical properties of the large scale structure of the Universe. Specifically, we consider here the gas pressure profile within galaxy clusters as a function of radius from the dark matter halo. Having a three point correlation between the lensing and SZ effect can constrain parameters in the pressure profile model to reveal new physics regarding the relationship between dark matter and gas pressure. 

The gas pressure profile of galaxy clusters and its relation to the SZ effect has been studied by several groups including \citet{Arnaud2010} and \citet{Planck2013}. In \citet{Planck2013} the reconstructed SZ map was used to study the pressure of 62 massive clusters. By stacking radial profiles, the gas pressure profile was measured and the best fit parameters were found to be $[P_0, c_{500},\alpha,\beta,\gamma] = [6.41,1.81,1.33,4.13,0.31]$.  Where $P_0, c_{500}, \alpha, \beta$ and $\gamma$ are the central pressure, concentration, central slope, intermediate slope and outer slope of dark matter halos, respectively. It was found that at large radii the pressure profile was flatter than simulations would predict. In \citet{Arnaud2010}, simulations and observations of 33 clusters from XMM-Newton are used to create a generalized Navarro-Frenk-White (NFW) \citep{Navarro1996} profile for gas pressure and the resulting parameters are  $[P_0, c_{500},\alpha,\beta,\gamma] = [8.40,1.18,1.05,5.49,0.308]$. While \cite{Arnaud2010} combined observation and simulation there remains tension between the two approaches. The CMB bispectrum measurement can be a complementary way to constrain the gas pressure parameters and provide observational evidence for understanding the tension between simulation and observation.

The paper is organized as follows: in the next section we discuss the skewness estimator and its derivation. In section 3 we review the data analysis preformed. In section 4 the results of the analysis are presented. In section 5 the MCMC analysis is discussed along with the results and their cosmological implications. Section 6 is a summary of the findings and suggestions for future work. 
Throughout we make use of the standard flat-$\Lambda$CDM cosmological model with $H_0$= 70 km s$^{-1}$ Mpc$^{-1}$ and $\Omega_{\Lambda}$=0.73.

\section{Estimator}
The derivation of the bispectrum and the skewness spectrum are discussed at length in several papers including \citet{Cooray2001} and more recently \citet{Munshi2011}. Here the authors cover the key points for this analysis and refer the reader to the previous work for a more detailed discussion. The angular bispectrum $B^{TT y}_{\ell_1\ell_2\ell_3}$ is defined as a triangle with sides ($\ell_1$,$\ell_2$,$\ell_3$) in multipole space where $T(\textbf{n})T(\textbf{n})$ and $y(\textbf{n})$ are statistically isotropic fields. With $T$ representing a temperature map and $y$ representing the SZ {\it y}-component map. The bispectrum is related to the multipole moments of the fields by:

\begin{equation}
B^{TTy}_{\ell_1\ell_2\ell_3} = \displaystyle\sum_{m_1m_2m_3}
\begin{pmatrix}
\ell_1 & \ell_2 & \ell_3 \\
m_1 & m_2 & m_2
\end{pmatrix} \\
\langle a^T_{\ell_1m_1} a^T_{\ell_2m_2}a^y_{\ell_3m_3} \rangle
\end{equation}

The skewness power spectrum is the correlation of the product map $T(\textbf{n})T(\textbf{n})$ and $y(\textbf{n})$. This is useful because the angular bispectrum $B^{TT y}_{\ell_1 \ell_2 \ell_3}$ can be difficult to measure fully. The skewness spectrum $C^{TT,y}_{\ell}$ is a summation of the triangular configurations keeping one of the sides length $\ell$ fixed. Following the discussion in \citealt{Cooray2001,Munshi2011,Calabrese2010} the bispectrum can be described by:

\begin{center}
\begin{eqnarray}
B^{TTy}_{\ell_1\ell_2\ell_3} &=& -[ C_{\ell}^{\phi y} C_{\ell_1} \frac{\ell_2(\ell_2 +1)-\ell_1(\ell_1 + 1 ) - \ell_3(\ell_3 +1)}{2} \nonumber\\
&+& {\rm perms.}]  \sqrt{\frac{(2\ell_1+1)(2\ell_2+1)(2\ell_3+1)}{4\pi}}
\begin{pmatrix}
\ell_1 & \ell_2 & \ell_3 \\
0 & 0 & 0
\end{pmatrix}\nonumber\\
\end{eqnarray}
\end{center}

Here $C_{\ell}^{\phi y}$ is the amplitude of the cross-correlation power spectrum between the lensing potential and the {\it y}-parameter map and $C_{\ell_1}$ is the unlensed CMB anisotropy power spectrum. Only the permutations in which $\ell_1$ and $\ell_2$ vary are used since $\ell_3$ remains tied to the secondary anisotropy and is fixed to relate to the skewness spectrum.

From \citealt{Munshi2011,Calabrese2010} the optimized skewness estimator begins with defining a set of nine weighted temperature maps:
\begin{eqnarray}
X^1_{\ell m} = \frac{a_{\ell m}}{\tilde{C_{\ell}}}C_{\ell}; Y^1_{\ell m} = \ell(\ell+1)\frac{a_{\ell m}}{\tilde{C_{\ell}}}; Z^1_{\ell m} = \frac{a_{\ell m}}{\tilde{C_{\ell}}}C_{\ell}^{\phi y}\nonumber\\
X^2_{\ell m} = -\ell(\ell+1)\frac{a_{\ell m}}{\tilde{C_{\ell}}}C_{\ell}; Y^2_{\ell m} = \frac{a_{\ell m}}{\tilde{C_{\ell}}}; Z^2_{\ell m} = \frac{a_{\ell m}}{\tilde{C_{\ell}}}C_{\ell}^{\phi y}\nonumber\\
X^3_{\ell m} = \frac{a_{\ell m}}{\tilde{C_{\ell}}}C_{\ell}; Y^3_{\ell m} = \frac{a_{\ell m}}{\tilde{C_{\ell}}}; Z^3_{\ell m} = -\ell(\ell+1)\frac{a_{\ell m}}{\tilde{C_{\ell}}}C_{\ell}^{\phi y}\nonumber\\
\end{eqnarray}

Where $\tilde{C_{\ell}} = C_{\ell}+N_{\ell}/b^2_{\ell}$ is the temperature power spectrum, $b_{\ell}$ is the beam transfer function and
$N_{\ell}$ is the noise power spectrum. The nine weighted maps are generated by $T^{(i)}(\textbf{n}) = \sum Y_{\ell m}(\textbf{n})T^{(i)}_{\ell m}$ where \textit{i} is the index of the weighted map from above. The optimized skew spectrum is defined as:

\begin{equation}
C^{XY,Z}_{\ell} = \frac{1}{2\ell+1}\displaystyle\sum_{i}\displaystyle\sum_{m}\rm{Real}[(X^{(i)}(\textbf{n})Y^{(i)}(\textbf{n}))_{\ell m}Z^{(i)}(\textbf{n})_{\ell m}]
\end{equation}

The measured skewness spectrum $\hat{C}^{TT,y}_{\ell}$ now can be related to the bispectrum as shown in \citealt{Cooray2001,Munshi2011} as:

\begin{equation}
\hat{C}^{XY,Z}_{\ell} = \frac{1}{2\ell+1}\displaystyle\sum_{\ell_1 \ell_2}\frac{\hat{B}_{\ell \ell_1 \ell_2}B_{\ell \ell_1 \ell_2}}{\tilde{C_\ell}\tilde{C_{\ell_1}}\tilde{C_{\ell_2}}}\label{est}
\end{equation}

Here $\hat{B}_{\ell \ell_1 \ell_2}$ is the reduced bispectrum, meaning it has been weighted as the maps have in the derivation of $C^{TT,y}_{\ell}$ and the calculation only includes the permutations in which $\ell_{3}$ is fixed. The range of $\ell$ is $2 <\ell< 1600$. 

Now the measured skewness spectrum $\hat{C}^{TT,y}_{\ell}$ can be related to the theoretical ${C}^{TT,y}_{\ell}$. The theory ${C}^{TT,y}_{\ell}$ is calculated analytically by plugging the bispectrum formulation in Equation 2 into Equation 5. Up to this point the amplitude of the lensing SZ cross-correlation $C_{\ell}^{\phi y}$ has been taken to be unity. While the measured spectrum contains one factor of $C_{\ell}^{\phi y}$ and one factor of $\hat{C}_{\ell}^{\phi y}$, the theory spectrum contains two factors of $C_{\ell}^{\phi y}$.  The ratio of the measured and theoretical spectra gives the measured lensing SZ cross-correlation $\hat{C}_{\ell}^{\phi y}$.

\section{Data Analysis}

For the purposes of this analysis the {\it Planck} PR2-2015 all sky maps were used. Specifically the 100 GHz, 143G Hz and 217 GHz temperature maps were used as well as the MILCA full mission \textit{y}-map component foreground map. The data were reduced using custom python scripts within the HEALPY\footnote{http://healpix.sourceforge.net} \citep{Gorski2005} code framework. Briefly, the temperature maps were masked using a combination of Galactic foreground mask and the {\it Planck} released point source map. The 60\% foreground mask in conjunction with a point source mask was utilized in order to mask out any contamination by the Milky Way galaxy and bright sources. The monopole signal as well as the dipole signal were modeled using HEALPY and removed before measuring the power spectrum. 

In order to measure the noise $N_\ell$ for the temperature maps 100 {\it Planck} released simulated noise maps were passed through the analysis pipeline and the resulting median power spectrum was determined to be the noise contribution to the  measured power spectrum. The CMB anisotropy power spectrum $C_\ell$ was generated using the CAMB \citep{Lewis2013} code, the result of which was in agreement with the measured CMB power spectrum released by the {\it Planck} team. To model the noise in the \textit{y}-map, the half difference of the first and last maps were used. 

To measure the direct cross-correlation between lensing and tSZ \citep{Spergel1999}, the {\it Planck} released lensing map was similarly masked before being cross correlated with the component \textit{y}-map to measure $C_{\ell}^{\phi y}$. To measure the dust contamination the {\it Planck} released dust map was used and subtracted from the temperature maps before being run through the data pipeline.

\begin{figure}
 \includegraphics[scale=.52]{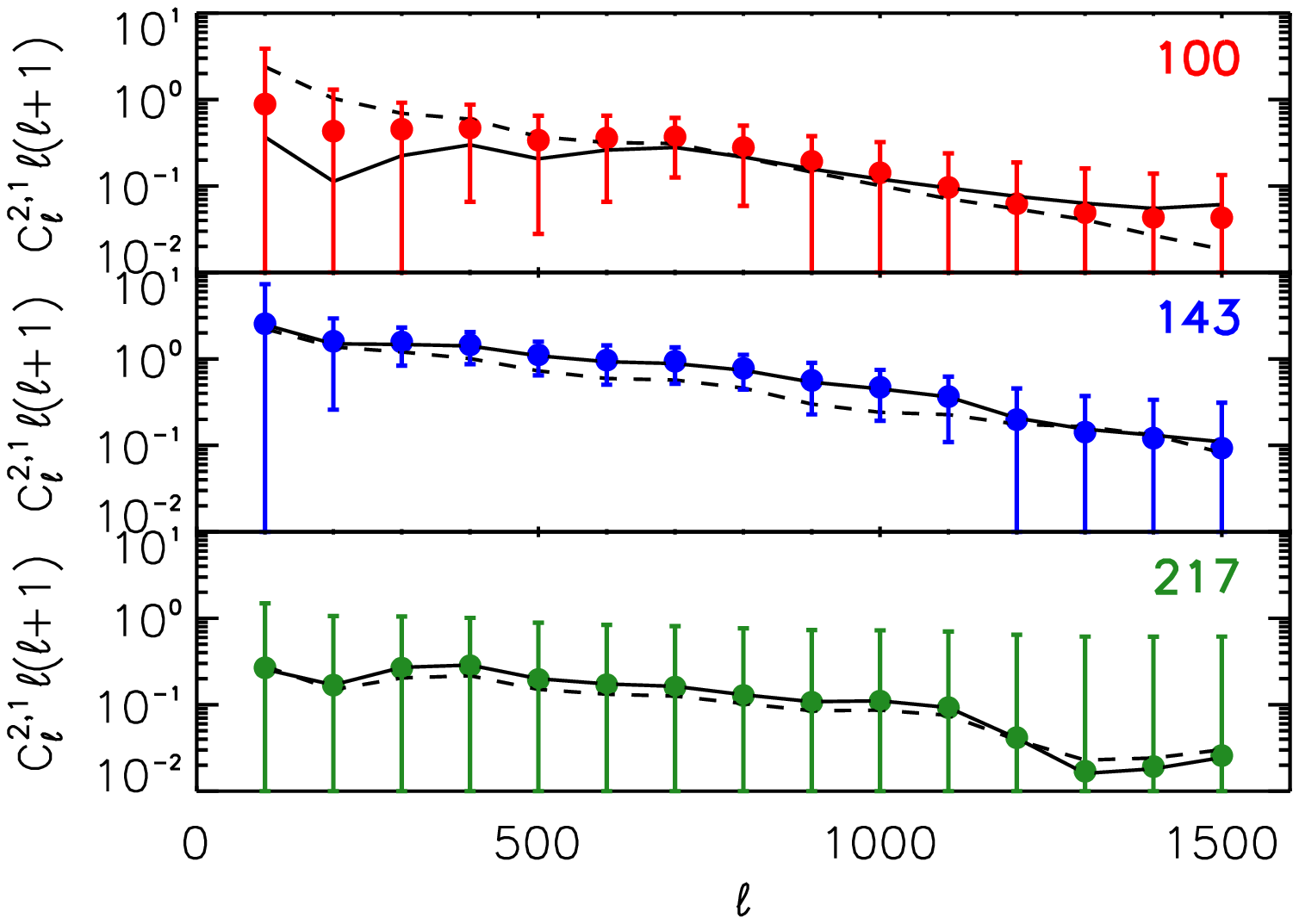}
  \includegraphics[scale=.52]{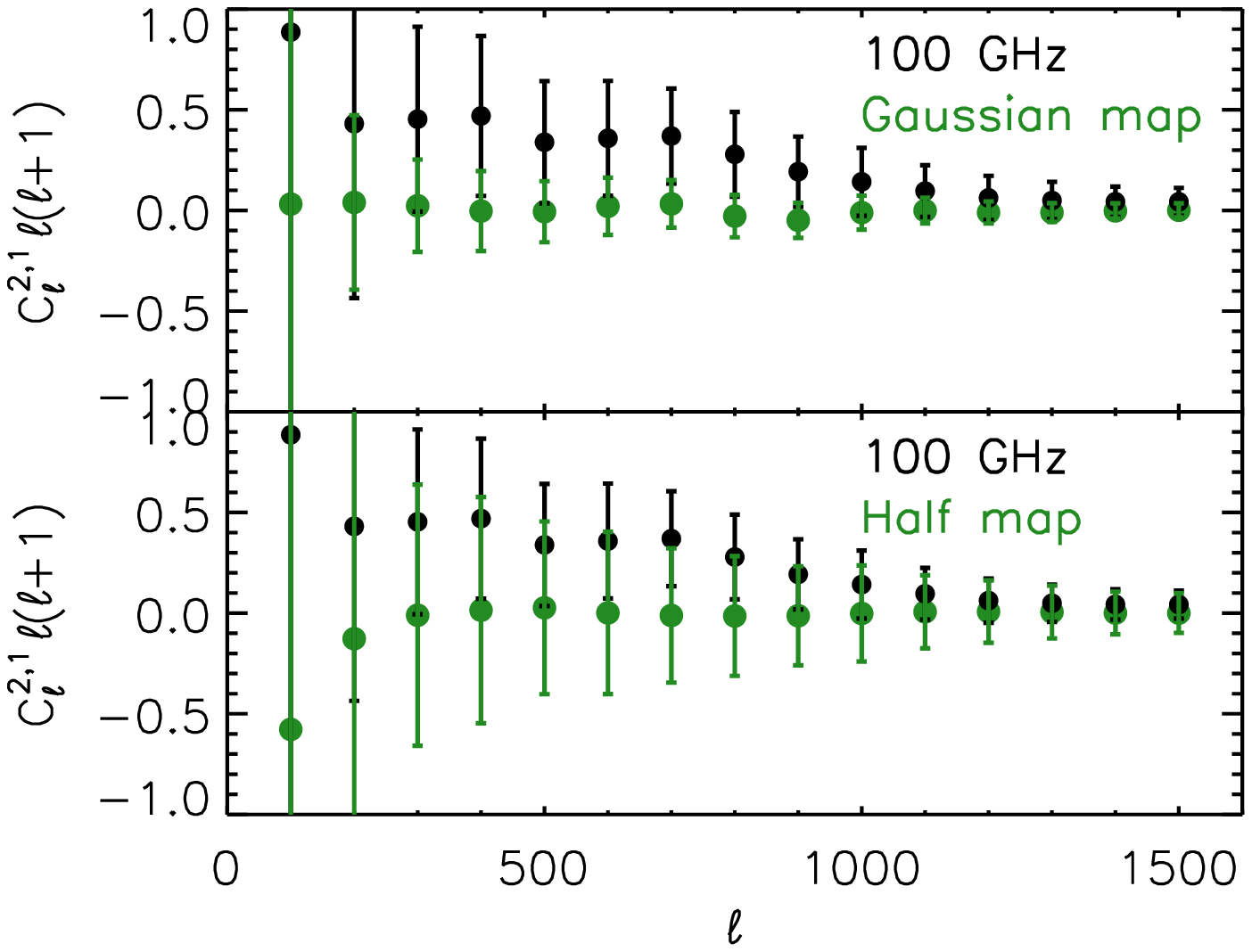}
  \caption {{\it top}: The measured skewness spectrum for each of the three frequencies. 100 GHz, 143GHz and 217 GHz from top to bottom respectively. The contributions of the dust signal and the SZ signal in the temperature maps are plotted as well. The dashed line is the resulting spectrum after subtracting the {\it Planck} dust map from the temperature map. The solid line is the resulting spectrum after having subtracted the frequency corrected \textit{y}-map from the temperature map. {\it bottom}: The result of passing a Gaussian map with noise through the estimator and the result of passing the half-map difference through the estimator. For comparison, the 100 GHz skewness spectrum is plotted. Both results are consistent with a null result as would be expected.}
\end{figure}

\section{Results}

In the top portion of Figure 1 we show the resulting skewness power spectra from the different frequencies with error bars from the simulated noise maps as well as cosmic variance. While there is a similar signal in the 100 and 143 GHz spectra the 217 GHz spectrum shows a deficiency in the SZ effect which is frequency dependent. Also plotted are the contribution of the dust and SZ effect on the temperature maps. The dust was subtracted from the temperature map and then the results were passed through the data pipeline which is displayed as the dashed line in the figure. The SZ map was also subtracted from the temperature maps and the result is plotted as a solid line. Neither the dust nor the SZ maps had a statistically important effect on the results in the 143 GHz and 217 GHz spectra but the 100 GHz map had a significant amount of signal removed by subtracting out the SZ map before the final measurement.

The results of the data pipeline null test are shown in the bottom portion of Figure 1. The upper part shows the result when passing a simple Gaussian map with simulated noise through the pipeline. The bottom part shows the result when using a map made up of the difference between two half-maps. For comparison, the 100 GHz spectrum is plotted. If there is no systematic contribution to the signal in the data pipeline both the Gaussian and half-map spectra should be consistent with zero. While the error is large in the half-map spectrum, both results show that the signal in the 100 GHz spectrum is not coming from systematics and is non-vanishing.

\begin{figure}
 \includegraphics[scale=.52]{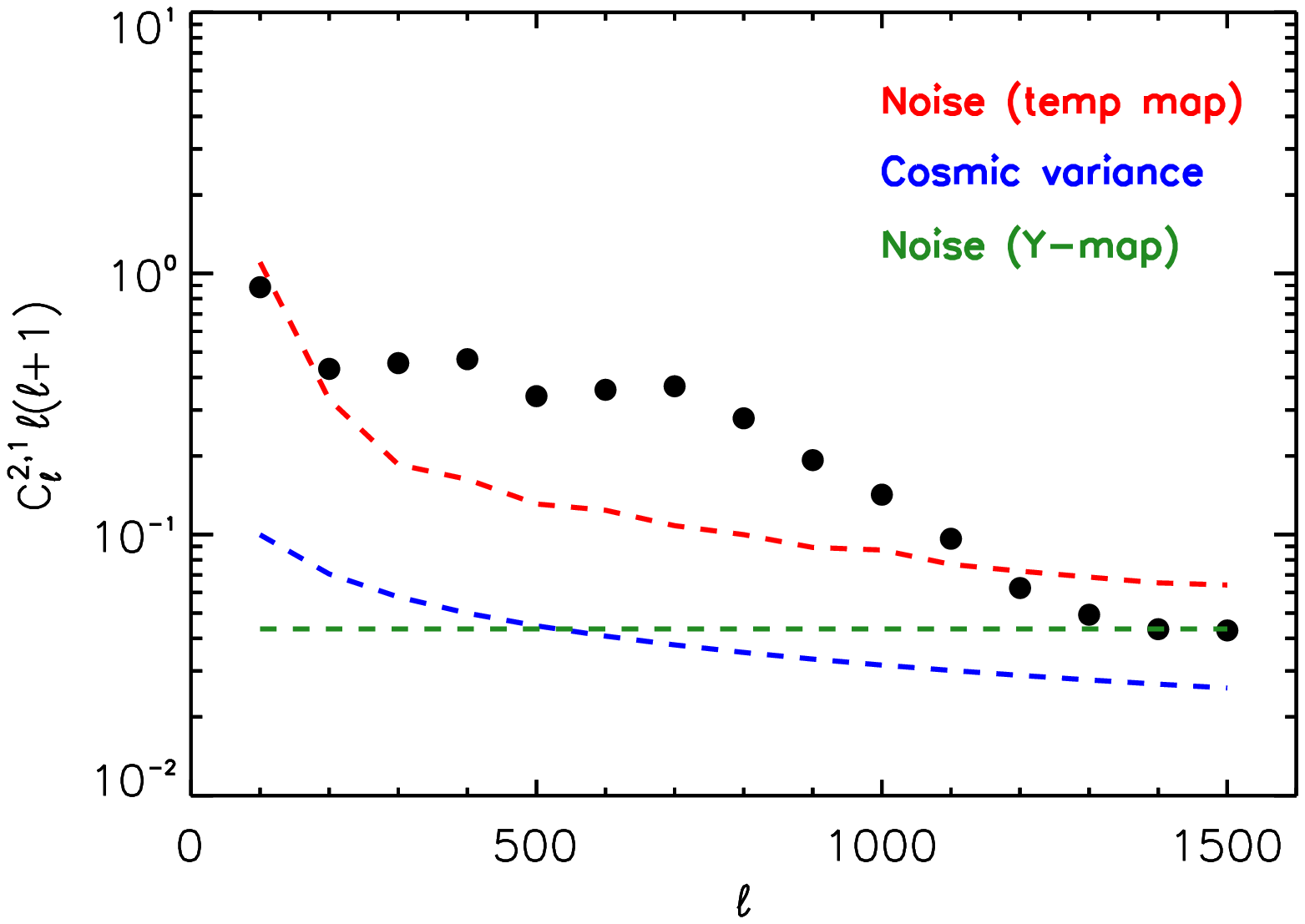}
  \includegraphics[scale=.52]{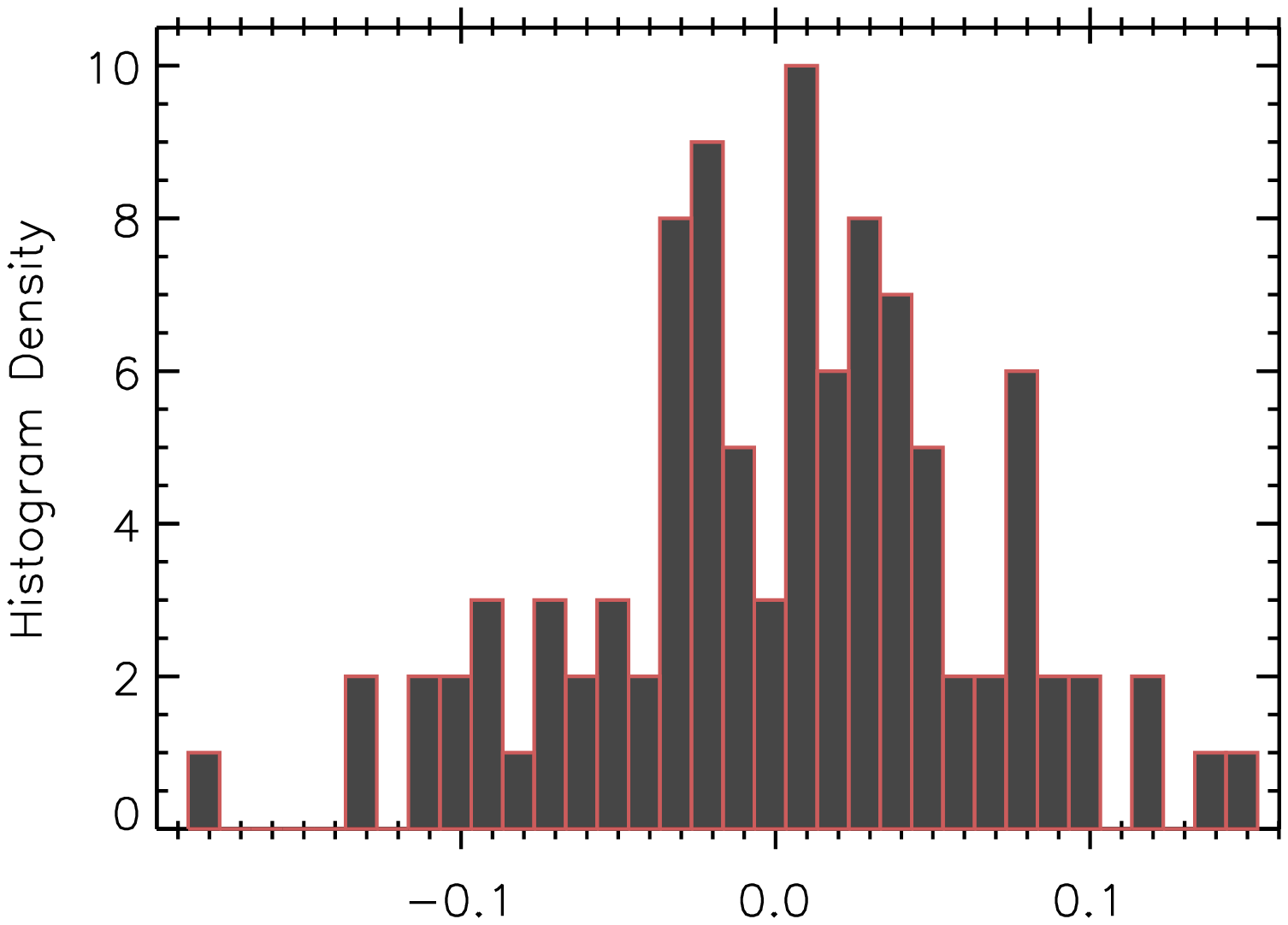}
  \caption {{\it top}: The component contributions to the total error. The 100 GHz spectrum is shown as an example which is representative of all three frequencies. {\it bottom}: A histogram of the variance in the signal of the simulated noise. 100 simulated maps were put through the same estimator as the data and the value at $\ell = 1000$ is plotted here. }
\end{figure}

The contributions from various error estimates are plotted in the top portion Figure 2. In the figure the 100 GHz skewness spectrum is plotted and is representative of the other frequencies. To estimate the error contribution from the temperature maps, 100 {\it Planck} simulated noise maps at each frequency were passed through the data pipeline. The standard deviation of the resulting power spectra became the error estimate. The dominate source of error is from the simulated temperature noise maps while the noise contribution from the \textit{y}-parameter map is not as significant.  The contribution from cosmic variance is not significant. At high $\ell$ all of the noise contributions become significant with the temperature map noise rising above the signal. The bottom portion of Figure 2 contains a histogram plot of the skew spectrum value at $\ell = 1000$ for the 100 GHz simulated noise maps is shown. The noise contribution is almost Gaussian as expected over 100 simulated maps.

\section{MCMC and Model Interpretation}

\begin{figure}
 \includegraphics[width=8cm,height=6cm]{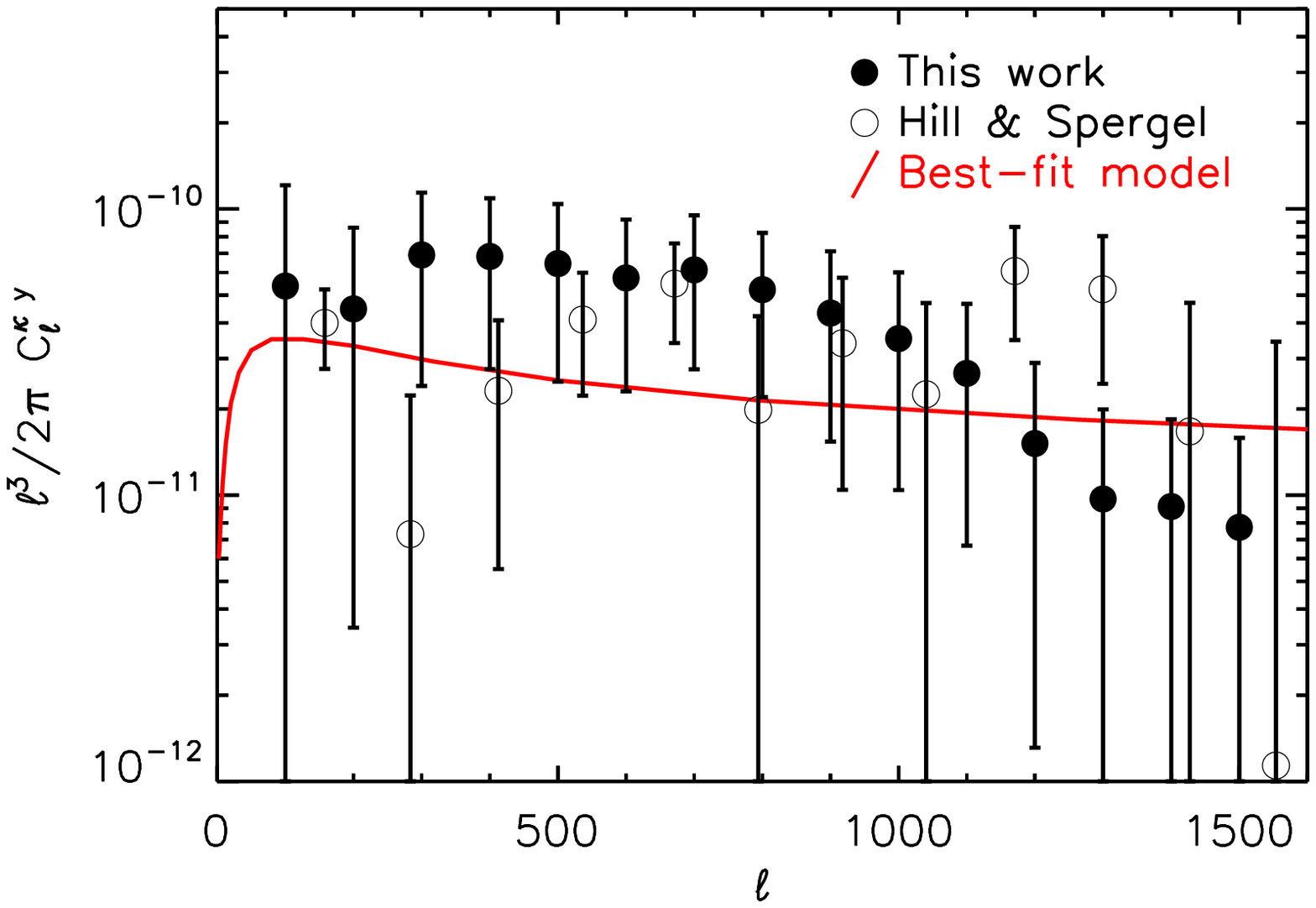}
 \includegraphics[width=8cm,height=6cm]{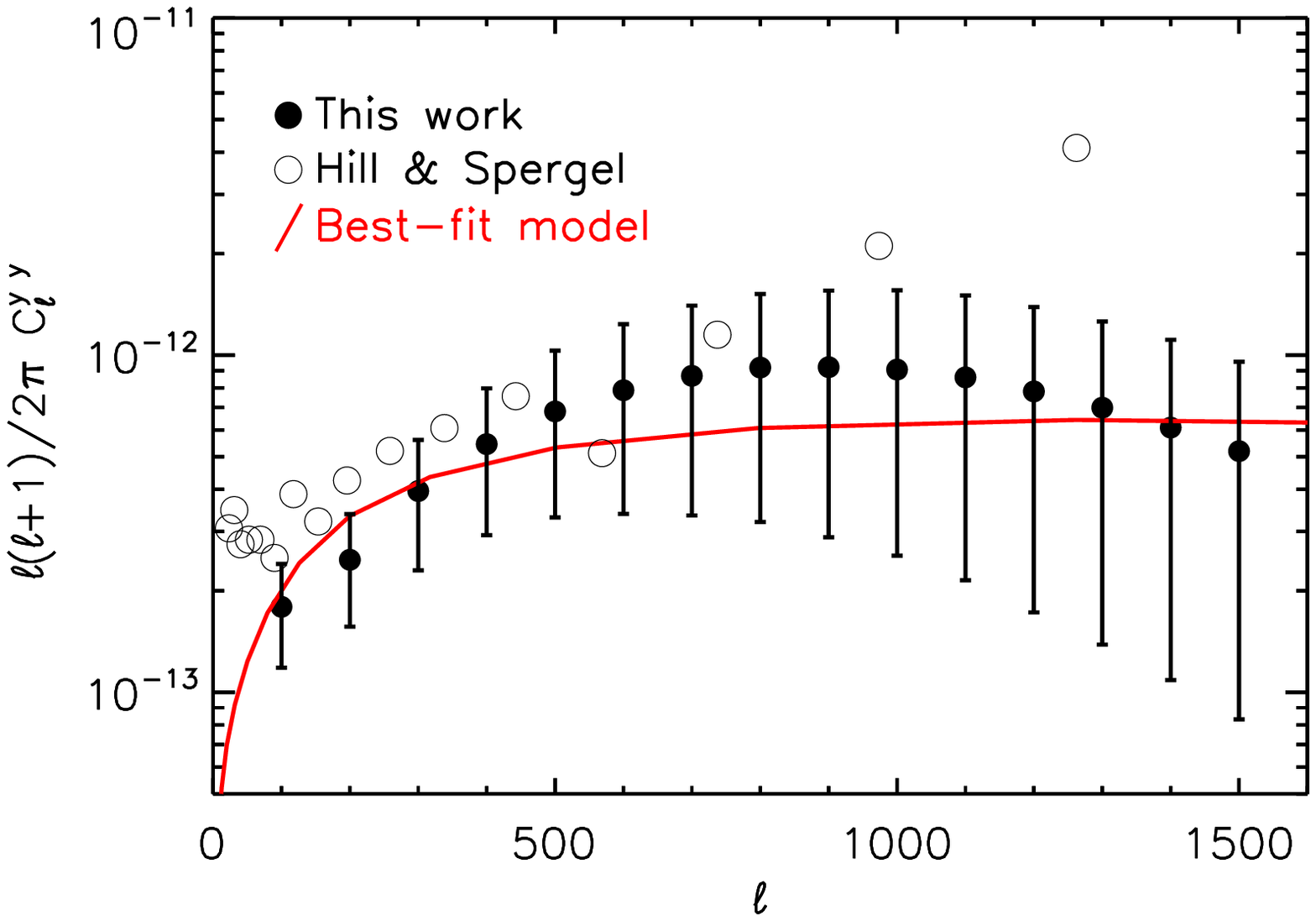}
  \caption { The best fit models the cross power spectrum $C_{\ell}^{\kappa y}$ and auto power spectrum $C_{\ell}^{yy}$ are plotted in red. They are compared to the measured cross power spectrum from the skew spectrum analysis as well as the auto power spectrum of the {\it Planck} \textit{y}-parameter maps. The results from \citet{Hill2014} are shown as a comparison, being the direct two point lensing-SZ cross correlation and the \textit{y}-map auto spectrum from the generated maps therein.}
\end{figure}

\begin{figure}
 \includegraphics[width=8cm,height=6cm]{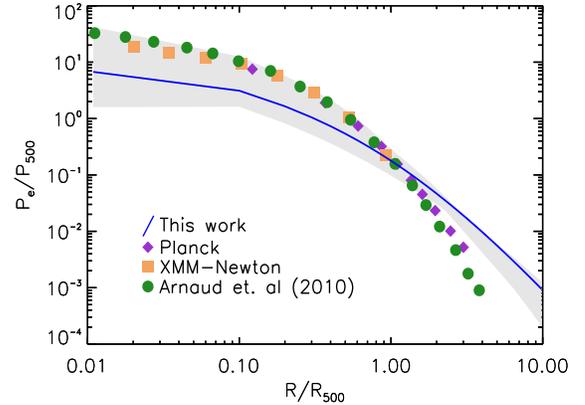}
  \caption { The best fit model of the gas pressure $P_e(r)$ including the 1$\sigma$ confidence region shown by the shaded region. For comparison the pressure profiles measured in \citealt{Planck2013, Arnaud2010} are also shown. }
\end{figure}

Figure 3 shows the measured three point lensing convergence-SZ cross-correlation power spectrum  $C_{\ell}^{\kappa y}$ and the measured two point \textit{y}-parameter auto power spectrum $C_{\ell}^{yy}$. From the Monte Carlo Markov Chain (MCMC) procedure, we obtain the best fit models for each case and have plotted the result. The measured lensing-SZ cross-correlation comes from the resulting skewness power spectra which has been averaged over the three frequency bands with error added in quadrature. Also shown is the two point lensing-SZ cross correlation measured in \citet{Hill2014}. The two point \textit{y}-parameter auto power spectrum is measured by taking the auto-spectrum of the {\it Planck} \textit{y}-parameter map. The points from \citet{Hill2014} are again shown for comparison.

The best fit model for the gas pressure from the three-point correlation is shown in Figure 4 along with the 1$\sigma$ confidence region. For comparison gas pressure profiles from {\it Planck} and XMM-{\it Newton} are shown \citep{Planck2013, Arnaud2010}. 

We follow the gas pressure model in \citet{Komatsu2011}, i.e.,
\begin{eqnarray}
P_e(r)&=&1.65h_{70}^2E^{8/3}(z)[\frac{M_{500}}{3\times 10^{14}M_{\odot}/h_{70}}]^{2/3+\alpha_P}\nonumber\\
&\times&\tilde P(x)\,\,[\rm{eV}/\rm{cm}^3].\label{pr}
\end{eqnarray}

Here $E(z)=H(z)/H_0$, $h_{70}=h/0.7$ and $\alpha_P=0.12$. We determine the radius $r_{500}$ from the relation $M_{500}=4\pi/3[500\rho_c(z)]r^3_{500}$. In this equation, $\rho_c(z)$ is the critical density and $x=r/r_{500}$.

The radial part of the pressure $\tilde P(x)$ in Equation~\ref{pr} is parametrized as
\begin{equation}
\tilde P(x)=\frac{P_0}{(c_{500}x)^{\gamma}[1+(c_{500}x)^{\alpha}]^{(\beta-\gamma)/\alpha}},
\end{equation}
which is different from the well-known NFW dark matter halo profile because the gas is a biased tracer of DM. The gas pressure profile is fully determined by the parameter set $\{P_0,c_{500},\alpha,\beta,\gamma\}$.

We use the halo model \citep{Cooray2002} to predict the theoretical power spectra $C_{\ell}^{yy}$ and $C_{\ell}^{\kappa y}$ following the calculations of 1-halo and 2-halo terms \citep{Planck2014, Hill2014,Battaglia2015}. We then compare the theoretical power spectra to the measured ones, sampling the gas pressure profiles and generating posterior distribution functions from the following likelihood
\begin{equation}
-2\ln \mathcal{L}=\displaystyle\sum_{i=\kappa y, yy}\displaystyle\sum_b\Big[\frac{C^{\rm{obs},(i)}_{b}-\hat C_{b}^{(i)}}{\Delta_b}\Big]^2 + \rm{const}.
\end{equation}

\section{Discussion}
As demonstrated, the three-point measurement from the skew spectrum is in agreement with the two-point direct measurement from the literature. It serves as an independent check on the work done before and can be used to constrain physical properties like the gas pressure profile. The non-zero result confirms the correlation between dark matter CMB lensing and the hot ICM traced by the thermal SZ effect.

Using only the CMB, the gas pressure profile is examined as a separate check on the work done by observing clusters directly or by simulation. The resulting gas pressure profile from the MCMC procedure is defined by the parameters $\{P_0,c_{500},\alpha,\beta,\gamma\}$ are \{$5.6^{+2.2}_{-2.2}$, $2.0^{+0.5}_{-0.5}$, $1.4^{+0.4}_{-0.4}$, $3.3^{+0.7}_{-0.4}$, $0.4^{+0.1}_{-0.2}$\}. These parameters are in agreement with the literature in which observations and simulations are conducted to measure the profile. It should be noted that the work done here assumed a z = 0 and will hopefully be able to be compared to high z clusters in future work in order to better understand the evolution of the gas pressure profile.

 Interestingly, the parameter $\beta = 3.3^{+0.7}_{-0.4}$ in this work is comparatively overestimated, in \citep{Planck2013} $\beta = 4.13$  and in \citep{Arnaud2010} $\beta$ = 5.49. The parameter $\beta$ is the outer slope of the pressure profile and the lower value corresponds to a greater thermal pressure at large radius. More than 50\% of the thermal SZ signal comes from $R > R_{500}$ and should be sensitive to the power spectrum measured herein \citep{Ramos2015}. 
 
 As can be seen in Figure 4 the pressure profile from this work along with {\it Planck} and XMM-{\it Newton} are in agreement besides the divergence at large R. As a check the two point correlation was also modeled for the pressure profile and found to not vary significantly from the three-point model.

\section{Summary}

The CMB skewness power spectrum was measured using the correlation between {\it Planck} frequency maps and {\it Planck} Sunyaev-Zel'dovich (SZ) component maps. We also measure the lensing-thermal SZ cross-correlation power spectrum using the three-point correlation function (bispectrum) and compare it to the two-point correlation direct measurements from {\it Planck} all sky maps.

 The Markov Chain Monte Carlo (MCMC) procedure is utilized in order to find the best fit lensing cross-correlation power spectrum as well as the best fit \textit{y}-parameter auto-spectrum.  The best fit and uncertainty values for the gas pressure parameters are found to be $\{P_0,c_{500},\alpha,\beta,\gamma\}$ are \{$5.6^{+2.2}_{-2.2}$, $2.0^{+0.5}_{-0.5}$, $1.4^{+0.4}_{-0.4}$, $3.3^{+0.7}_{-0.4}$, $0.4^{+0.1}_{-0.2}$\} Where $P_0, c_{500}, \alpha, \beta$ and $\gamma$ are the central pressure, concentration, central slope, intermediate slope and outer slope respectively. The parameters are found to be in agreement with literature.

\acknowledgments
NT, CF and AC would like to acknowledge support from NASA grants NASA NNX16AJ69G, NASA NNX16AF39G, 
HST-AR-13886.001-A, IGPP LANL 368641, NSF AST-1313319 and Ax Foundation for Cosmology at UC San Diego.
Some of the results in this paper have been derived using the HEALPix (K.M. Gorski et al., 2005, ApJ, 622, p759) package.


\newpage

\begin{thebibliography}{}
\expandafter\ifx\csname natexlab\endcsname\relax\def\natexlab#1{#1}\fi

\bibitem[{{Arnaud} {et~al.}(2010){Arnaud}, {Pratt}, {Piffaretti},
  {B{\"o}hringer}, {Croston}, \& {Pointecouteau}}]{Arnaud2010}
{Arnaud}, M., {Pratt}, G.~W., {Piffaretti}, R., {et~al.} 2010, \aap, 517, A92

\bibitem[{{Battaglia} {et~al.}(2015){Battaglia}, {Hill}, \&
  {Murray}}]{Battaglia2015}
{Battaglia}, N., {Hill}, J.~C., \& {Murray}, N. 2015, \apj, 812, 154

\bibitem[{{Calabrese} {et~al.}(2010){Calabrese}, {Smidt}, {Amblard}, {Cooray},
  {Melchiorri}, {Serra}, {Heavens}, \& {Munshi}}]{Calabrese2010}
{Calabrese}, E., {Smidt}, J., {Amblard}, A., {et~al.} 2010, \prd, 81, 043529

\bibitem[{{Cooray}(2001)}]{Cooray2001}
{Cooray}, A. 2001, \prd, 64, 043516

\bibitem[{{Cooray} {et~al.}(2000){Cooray}, {Hu}, \& {Tegmark}}]{Cooray2000}
{Cooray}, A., {Hu}, W., \& {Tegmark}, M. 2000, \apj, 540, 1

\bibitem[{{Cooray} \& {Sheth}(2002)}]{Cooray2002}
{Cooray}, A., \& {Sheth}, R. 2002, \physrep, 372, 1

\bibitem[{{G{\'o}rski} {et~al.}(2005){G{\'o}rski}, {Hivon}, {Banday},
  {Wandelt}, {Hansen}, {Reinecke}, \& {Bartelmann}}]{Gorski2005}
{G{\'o}rski}, K.~M., {Hivon}, E., {Banday}, A.~J., {et~al.} 2005, \apj, 622,
  759

\bibitem[{{Hill} \& {Spergel}(2014)}]{Hill2014}
{Hill}, J.~C., \& {Spergel}, D.~N. 2014, \jcap, 2, 030

\bibitem[{{Hirata} {et~al.}(2008){Hirata}, {Ho}, {Padmanabhan}, {Seljak}, \&
  {Bahcall}}]{Hirata2008}
{Hirata}, C.~M., {Ho}, S., {Padmanabhan}, N., {Seljak}, U., \& {Bahcall}, N.~A.
  2008, \prd, 78, 043520

\bibitem[{{Komatsu} {et~al.}(2011){Komatsu}, {Smith}, {Dunkley}, {Bennett},
  {Gold}, {Hinshaw}, {Jarosik}, {Larson}, {Nolta}, {Page}, {Spergel},
  {Halpern}, {Hill}, {Kogut}, {Limon}, {Meyer}, {Odegard}, {Tucker}, {Weiland},
  {Wollack}, \& {Wright}}]{Komatsu2011}
{Komatsu}, E., {Smith}, K.~M., {Dunkley}, J., {et~al.} 2011, \apjs, 192, 18

\bibitem[{Lewis(2013)}]{Lewis2013}
Lewis, A. 2013, Phys. Rev., D87, 103529

\bibitem[{{Munshi} {et~al.}(2011){Munshi}, {Heavens}, {Cooray}, \&
  {Valageas}}]{Munshi2011}
{Munshi}, D., {Heavens}, A., {Cooray}, A., \& {Valageas}, P. 2011, \mnras, 414,
  3173

\bibitem[{{Navarro} {et~al.}(1996){Navarro}, {Frenk}, \& {White}}]{Navarro1996}
{Navarro}, J.~F., {Frenk}, C.~S., \& {White}, S.~D.~M. 1996, \apj, 462, 563

\bibitem[{{Okamoto} \& {Hu}(2003)}]{Okamoto2003}
{Okamoto}, T., \& {Hu}, W. 2003, \prd, 67, 083002

\bibitem[{{Planck Collaboration} {et~al.}(2011){Planck Collaboration}, {Ade},
  {Aghanim}, {Arnaud}, {Ashdown}, {Aumont}, {Baccigalupi}, {Baker}, {Balbi},
  {Banday}, \& et~al.}]{Planck2011}
{Planck Collaboration}, {Ade}, P.~A.~R., {Aghanim}, N., {et~al.} 2011, \aap,
  536, A1

\bibitem[{{Planck Collaboration} {et~al.}(2013){Planck Collaboration}, {Ade},
  {Aghanim}, {Arnaud}, {Ashdown}, {Atrio-Barandela}, {Aumont}, {Baccigalupi},
  {Balbi}, {Banday}, \& et~al.}]{Planck2013}
---. 2013, \aap, 550, A131

\bibitem[{{Planck Collaboration} {et~al.}(2014){Planck Collaboration}, {Ade},
  {Aghanim}, {Armitage-Caplan}, {Arnaud}, {Ashdown}, {Atrio-Barandela},
  {Aumont}, {Baccigalupi}, {Banday}, \& et~al.}]{Planck2014}
---. 2014, \aap, 571, A21

\bibitem[{{Ramos-Ceja} {et~al.}(2015){Ramos-Ceja}, {Basu}, {Pacaud}, \&
  {Bertoldi}}]{Ramos2015}
{Ramos-Ceja}, M.~E., {Basu}, K., {Pacaud}, F., \& {Bertoldi}, F. 2015, \aap,
  583, A111

\bibitem[{{Sachs} \& {Wolfe}(1967)}]{SW1967}
{Sachs}, R.~K., \& {Wolfe}, A.~M. 1967, \apj, 147, 73

\bibitem[{{Smith} {et~al.}(2007){Smith}, {Zahn}, \& {Dor{\'e}}}]{Smith2007}
{Smith}, K.~M., {Zahn}, O., \& {Dor{\'e}}, O. 2007, \prd, 76, 043510

\bibitem[{{Spergel} \& {Goldberg}(1999)}]{Spergel1999}
{Spergel}, D.~N., \& {Goldberg}, D.~M. 1999, \prd, 59, 103001

\bibitem[{{Sunyaev} \& {Zeldovich}(1980)}]{SZ1980}
{Sunyaev}, R.~A., \& {Zeldovich}, I.~B. 1980, \mnras, 190, 413

\end{thebibliography}
\end{document}